\documentclass[8pt]{article}
\usepackage[margin=20mm]{geometry}

\usepackage{cite}
\usepackage{amsmath,amssymb,amsfonts}
\usepackage[ruled]{algorithm2e}
\usepackage[dvipdfmx]{graphicx}
\usepackage{textcomp}
\usepackage{xcolor}
\usepackage{url}
\usepackage{comment}
\usepackage{booktabs}
\usepackage{tabularx}

\usepackage{IEEEtrantools}
\usepackage[driverfallback=dvipdfm]{hyperref}

\def\BibTeX{{\rm B\kern-.05em{\sc i\kern-.025em b}\kern-.08em
　　T\kern-.1667em\lower.7ex\hbox{E}\kern-.125emX}}
\begin{document}
\bstctlcite{IEEEexample:BSTcontrol}

\title{Machine Learning-assisted High-speed Combinatorial Optimization with Ising Machines for Dynamically Changing Problems} 

\author{
Yohei Hamakawa$^{1,\ast}$, Tomoya Kashimata$^{1}$, Masaya Yamasaki$^{1}$, Kosuke Tatsumura$^{1}$\\
\small $^{1}$Computer \& Network System R\&D Dept., AI Digital R\&D Center, Toshiba Corporation, Kawasaki 212-8582, Japan\\
\small $^{\ast}$Corresponding author: Yohei Hamakawa(e-mail: yohei.hamakawa@toshiba.co.jp)
}
\date{}

\maketitle

\begin{abstract}
Quantum or quantum-inspired Ising machines have recently shown promise in solving combinatorial optimization problems in a short time. Real-world applications, such as time division multiple access (TDMA) scheduling for wireless multi-hop networks and financial trading, require solving those problems sequentially where the size and characteristics change dynamically. However, using Ising machines involves challenges to shorten system-wide latency due to the transfer of large Ising model or the cloud access and to determine the parameters for each problem. Here we show a combinatorial optimization method using embedded Ising machines, which enables solving diverse problems at high speed without runtime parameter tuning. We customize the algorithm and circuit architecture of the simulated bifurcation-based Ising machine to compress the Ising model and accelerate computation and then built a machine learning model to estimate appropriate parameters using extensive training data. In TDMA scheduling for wireless multi-hop networks, our demonstration has shown that the sophisticated system can adapt to changes in the problem and showed that it has a speed advantage over conventional methods.
\end{abstract}

\section{Introduction}\label{sec:introduction} 

\begin{figure}[t]
\centering
\includegraphics[width=17.2 cm]{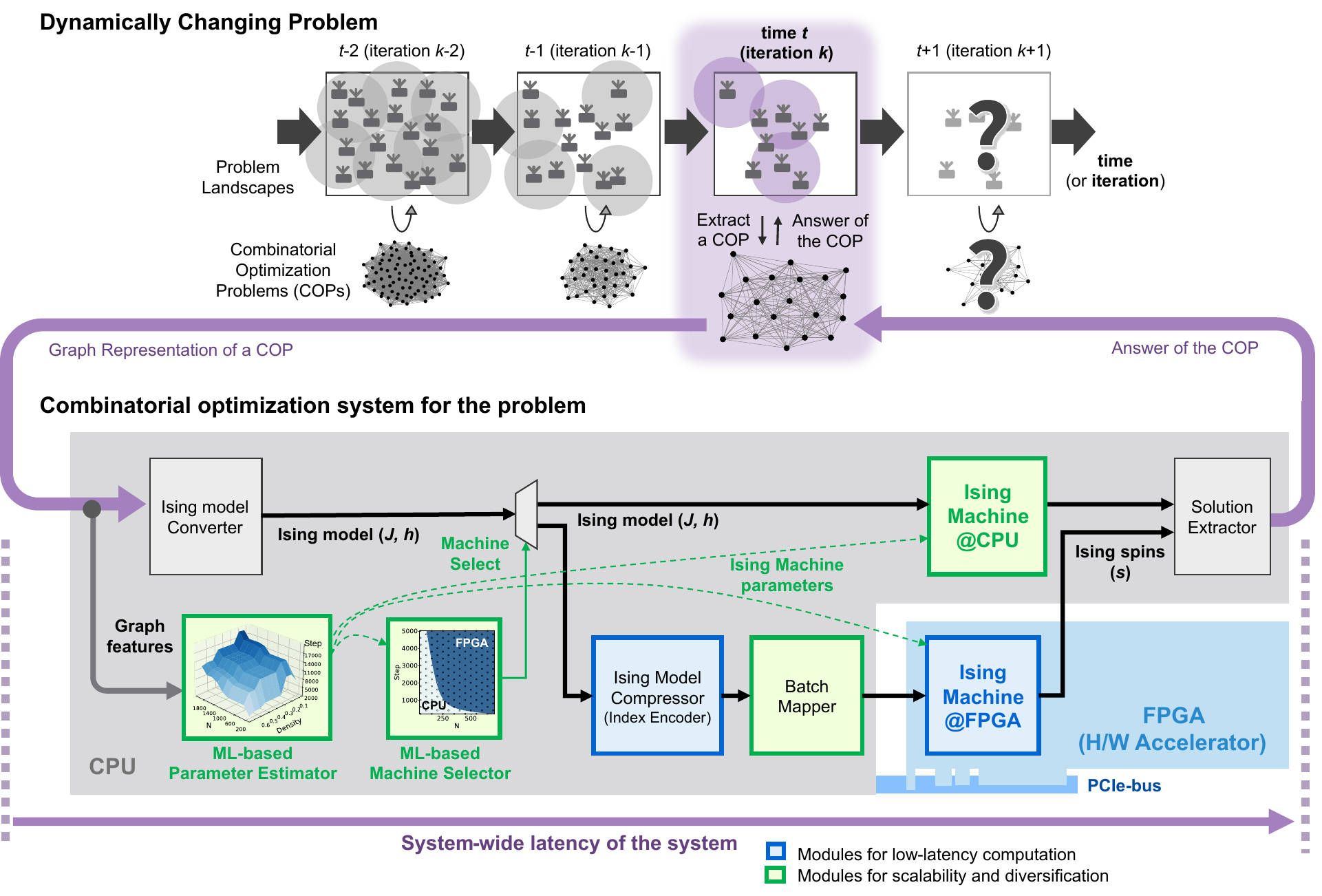}
\caption{A dynamically changing problem that include sequential combinatorial optimization and a combinatorial optimization system using embeddable Ising machines.  The system contains modules for low latency computation (blue-highlighted) and modules for scalability and diversification (green-highlighted) to handle the sequential problems having various sizes and characteristics.
}
\label{Fig_OverView}
\end{figure}
Combinatorial optimization is an important problem that frequently appears in various aspects of society and industry. Many combinatorial optimization problems are difficult to solve due to the large number of possible combinations as the number of variables increases. Ising machines are dedicated computers that can find approximate solutions to those combinatorial optimization problems at high speed and have recently attracted attention.

In some real-world applications like real-time processing, it is necessary to sequentially and rapidly solve multiple combinatorial optimization problems with different characteristics. Ising machines can be used for those problems. The examples of repetitive problems include time division multiple access (TDMA) scheduling for wireless multi-hop networks~\cite{ishizaki19} and clustering algorithms for unevenly distributed data~\cite{Matsumoto22}. In the former, Ising machines repeatedly solve the maximum independent set (MIS) problems that change every time to minimize the total number of TDMA slots. Real-time applications using Ising machines include wireless resource allocation~\cite{Hasegawa21}, financial trading~\cite{ISCAS20, ACCESS23a, ACCESS23b}, multiple object tracking for in-vehicle systems ~\cite{Tatsumura24}, traffic-congestion prediction~\cite{Pan23}, and advertisement recommendation~\cite{Mo20}.  In this article, we collectively refer to those as \textit{dynamically changing problems}~(the upper part of Fig.\ref{Fig_OverView}).  The dynamically changing problems cannot be solved in parallel because the optimization problem for the next (iteration)  cannot be determined until the current problem is solved. Therefore, it is necessary to solve each problem with a low latency at the system-wide level (to minimize the cumulative latencies). Some previous studies~\cite{ishizaki19,Matsumoto22,Hasegawa21,ISCAS20, ACCESS23a, ACCESS23b,Tatsumura24,Pan23,Mo20} have demonstrated the applicability of Ising machines to the dynamically changing problems under constant problem settings. However, generalized methods applicable to the cases where the sizes and charactersitics of individual problems varies greatly have not yet been proposed.

Ising machines are domain-specific computers designed to solve difficult combinatorial optimization problems in a short time~\cite{Finocchio24, sbm1, FPL19, sbm2, NatEle, kanao23,Kashimata24,Matsumoto22, johnson11,king23, honjo21, kalinin20, PoorCIM19, cai20, borders19, aadit22, litvinenko23, Graber24, moy22, albertsson21, wang21, sharma22, kawamura23, matsubara20, waidyasooriya21, okuyama19,SimCIM21}.  Ising machines search for the ground state of the energy of an Ising model~\cite{brush67}, which consists of binary variables called spins (\textbf{\textit{s}}) that interact with each other. The Ising problem is equivalent to quadratic unconstrained binary optimization (QUBO) and belongs to the nondeterministic polynomial time (NP)-hard class~\cite{barahona82, lucas14}; various types of computationally-hard combinatorial optimization problems can be formulated as Ising problems~\cite{lucas14}.  The Ising machines have been implemented with a variety of hardware~\cite{Finocchio24} including superconducting qubits~\cite{johnson11,king23}, optical systems~\cite{honjo21, kalinin20, PoorCIM19}, memristor-based neural networks~\cite{cai20}, probabilistic bits~\cite{borders19,aadit22}, spintronics systems~\cite{litvinenko23}, coupled oscillators~\cite{Graber24,moy22,albertsson21,wang21,sbm1}, analog computing units~\cite{sharma22}, application specific integrated circuits (ASICs)~\cite{kawamura23,matsubara20}, field programmable gate arrays (FPGAs)~\cite{sbm1, FPL19, sbm2, NatEle, Kashimata24, waidyasooriya21,SimCIM21}, and graphics processing units (GPUs)~\cite{sbm1, sbm2, okuyama19}.  Typically, an Ising machine as an accelerator forms a combinatorial optimization system together with a CPU responsible for general-purpose computation and control. In such configurations, data compression may be performed to reduce the data transfer time between the CPU and the accelerator and/or to speed up the computations by the accelerator. For example, in the field of embedded neural networks, various methods such as quantization and pruning for trained data have been proposed~\cite{Courbariaux15,Han16}. Similarly in the field of Ising machines, some compression methods and compressed representations of the coupling matrix ($J$-matrix) data have been proposed~\cite{Shimomai24}.

A combinatorial optimization system for dynamically changing problems requires not only the high-speed Ising machine itself but also its input and output to be fast in order to minimize the system-wide latency. Embedded Ising machines using ASICs, FPGAs, or GPUs are needed instead of cloud services which may have queue waiting times or network delays.  To handle large size problems, it is desirable to perform data compression with low information loss. Furthermore, to obtain high-quality solutions with an Ising machine, it is necessary to consider the parameters related to the formulation in the form of QUBO or Ising models and the control parameters of the Ising machine ~\cite{kuroki24}. Generally, the optimal values of these parameters vary for each problem~\cite{yarkoni18, gherardi24,Komiyama21}. Since searching for these optimal parameters takes time, it is not practical to search for them each time for individual dynamically changing problems.

Here we propose a combinatorial optimization method with low system-wide latency that can be applied to dynamically changing problems using embedded Ising machines~(the lower part of Fig.\ref{Fig_OverView}).  This method is based on two key ideas: one is a set of techniques for low-latency computation, and the other is a set of techniques for scalability and diversification. We demonstrate the performance and practicality of the proposed method for TDMA scheduling for wireless multi-hop networks.

The low-latency computation consists of two methods. First, we implement a quantum-inspired algorithm called simulated bifurcation (SB)~\cite{sbm1, FPL19, sbm2, NatEle, kanao23,Kashimata24} on an FPGA as an embedded Ising machine. The algorithm of SB was derived in 2019~\cite{sbm1} through classicizing a quantum-mechanical Hamiltonian describing a quantum adiabatic optimization method~\cite{qbm} and improved in 2021~\cite{sbm2}, where two branches of the bifurcation in each simulated oscillator correspond to two states of each Ising spin. The SB algorithm has extremely high parallelism, enabling high-speed operation with a large number of processing units on the FPGA.  Among the different SB variations, we adopt the ballistic SB (bSB)~\cite{sbm2}, which is good at obtaining high-quality solutions with a small number of time-evolution steps.  Second, we propose a lossless compression encoding method for the $J$-matrix called \textit{indexing fast-computation architecture}.  The $J$-matrix is a large dataset with $\mathcal{O}(N^2)$ elements (where \textit{N} is the number of decision variables \textbf{\textit{s}}). In many problems in the form of Ising model~\cite{lucas14}, the elements ($J_{i,j}$) need real-number representation but the number of possible values that each element takes is limited (limited variation). The indexing fast-computation architecture reduces data transfer time between the CPU and the Ising machine by the lossless encoding based on those limited variations. Furthermore, we propose a method to perform most of the MAC (multiply and accumulation) operations without decoding the encoded state to original values (an encoded computation). This reduces computational resources needed and enables high-speed computation.

The scalability and diversification consists of three methods. First, we construct a machine learning-based parameter estimation model (parameter estimator). Using randomly generated network graphs and the corresponding parameters that can yield the best solutions faster as training dataset, a regression model is trained offline. At runtime, the parameter estimator can select the appropriate Ising machine parameters for each problem without overhead.  Second, there is a batch mapping mechanism of the $J$-matrix to effectively utilize the fixed-size memory of the FPGA-implemented Ising machine, even for small problems. This method allows multiple approximate solutions to be obtained simultaneously by duplicating a small problem and mapping them to a large $J$ memory.  Third, the proposed method is equipped with multiple Ising machines with different implementations and there is a selection mechanism of Ising machines depending on each problem. When the problem size is small, the overhead time of offloading to an FPGA-implemented Ising machine can be non-negligible. In such a case, a software-implemented Ising machine running on the CPU is selected instead of the FPGA-based one.

We implement SB-based Ising machines on an FPGA and a CPU and then construct a solver targeting the MIS problem, a representative NP-hard combinatorial optimization problem. We apply the MIS solver to a TDMA scheduling system for wireless multi-hop networks, which needs to rapidly and sequentially solve MIS problems having varying sizes and characteristics over many iterations.  By observing the system's ability to adapt the parameters/Ising machines to the problem variations, we validate that the advanced system is capable of adapting to problem variations and demonstrate its advantages in speed compared to a conventional MIS solver.

\section{Results}\label{sec_results}
\subsection{Indexing fast-computation architecture}\label{sec_indexed_j}

\begin{figure}[t]
\centering
\includegraphics[width=17.2 cm]{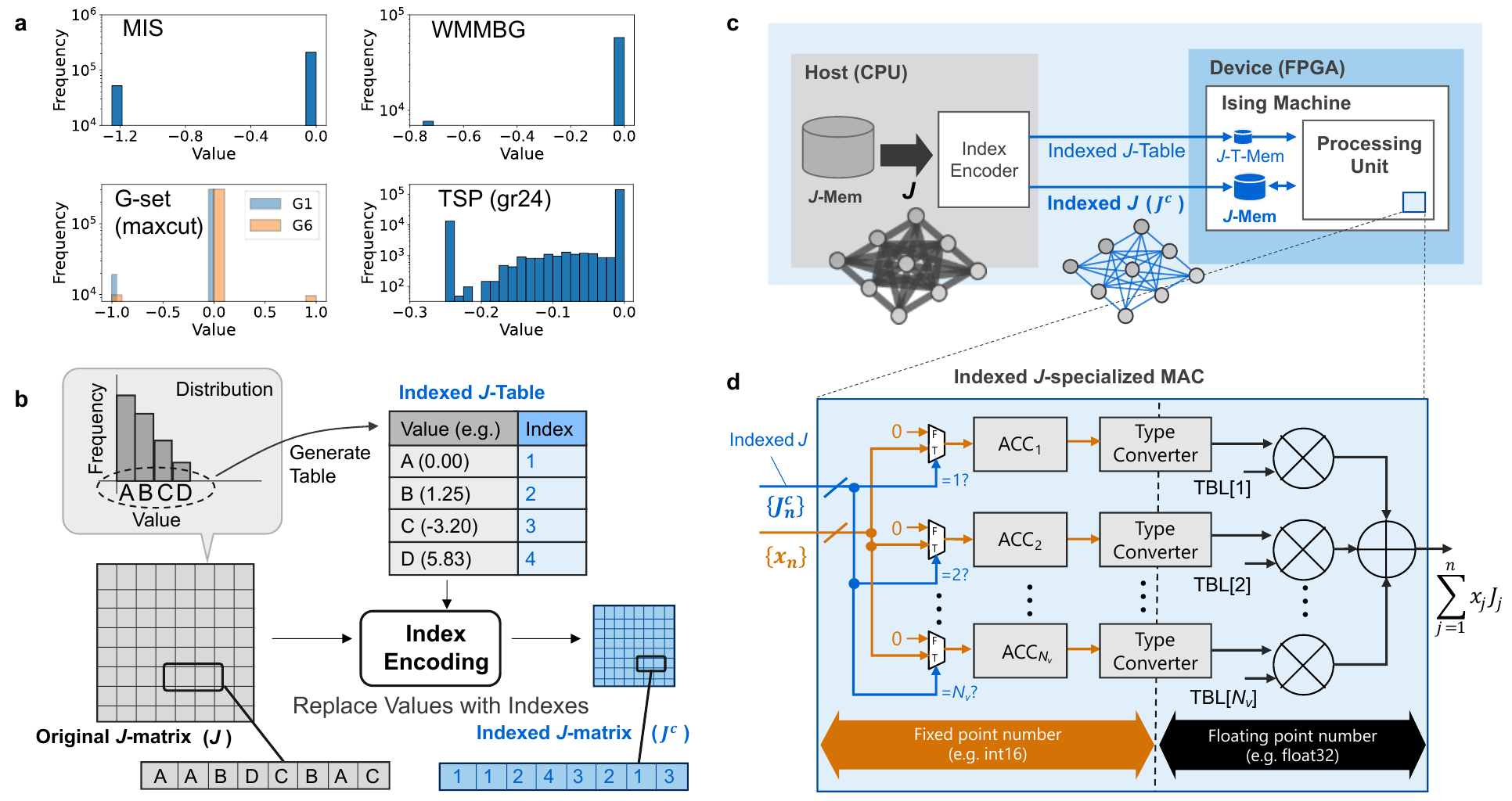}
\caption{Indexing fast-computation architecture.
(a) Distributions of $J$-matrix element values in Ising models, representing various combinatorial optimization problems.
(b) The method of indexing. An indexed $J$-Table is generated by extracting distinct values from the distribution of the $J$-matrix and then assign an index for each value.  By replacing the real values in the original $J$-matrix with the corresponding indices, a smaller-sized indexed $J$-matrix is produced.
(c) A system utilizing the encoding method. The indexed $J$ accelerates data transfer speed between the CPU and the Ising machine, and also reduces the memory requirements for the Ising machine.
(d) Indexed J-specialized MAC for simulated bifurcation (SB). By factorizing a MAC (multiply and accumulation) operation to isolate the additive part of the position variables \textit{x}, most computations can be performed without decoding the indexed \textit{J} from indices to real values, leading to faster computation.
}
\label{Fig_IFCA}
\end{figure}

Based on the characteristic data pattern of the discrete optimization problems formulated as the Ising model, the indexing fast-computing architecture performs data compression and high-speed computation.

The Ising problem is to find a spin configuration \textbf{\textit{s}} that minimizes the Hamiltonian (a cost function) $H_\mathrm{Ising}$ ~\cite{barahona82} defined by

\begin{equation}\label{IsingEnergy}
H_{\mathrm{Ising}}=-\frac{1}{2}\sum_{i=1}^{N}\sum_{j=1}^{N}J_{i,j}s_{i}s_{j}+\sum_{i=1}^{N}h_{i}s_{i},
\end{equation}
where $N$ is the number of spins, $s_{i}\;(\in \{-1,1\})$ is the $i$th Ising spin, $J_{i,j}$ is the coupling coefficient between the $i$th and $j$th spins, and $h_{i}$ is the bias coefficient (local field) for the $i$th spin.  Many combinatorial optimization problems can be formulated as Ising problems~\cite{lucas14}.  The MIS problem, which we focus on in this work, is to find the maximum independent set of a graph $G(V,E)$ with a node set $V$ and an edge set $E$. It can be  formulated as folows:

\begin{equation}\label{Eq_def_b}
s_i=
\begin{cases}
1, & (\text{Node } i \text{ is an element of the independent set}),\\
-1, & (\text{Node } i \text{ is not an element of the independent set}).
\end{cases}
\end{equation}
\begin{align}
J_{i,j} &= -\frac{A}{2}G_{i,j}, \label{Eq_mis_J}\\
h_i &= \frac{1}{2}\left( A\sum_{j=1}^N G_{i,j} - B\right) \label{Eq_mis_h}.
\end{align}
\begin{equation}\label{Eq_def_Gig}
G_{i,j}=
\begin{cases}
1, & (\text{Edge between } i \text{ and } j \text{ is an element of the edge set } E),\\
0, & (\text{Edge between } i \text{ and } j \text{ is not an element of the edge set } E).
\end{cases}
\end{equation}
Here, the coefficients \textit{A }and \textit{B} are positive real constants (see the Methods section for details). Unless there are special constraints, each element of \textit{J} and \textit{h} must be represented as real numbers.  When representing real numbers on a computer, single precision (32-bit) or double precision (64-bit) floating-point expressions are often used. In those cases, the data size of the $J$-matrix, which has $N^2$ elements, can become enormous, and floating-point arithmetic requires abundant computational resources.

The literature~\cite{lucas14} have provided Ising formulations for many NP-complete and NP-hard problems. In many of those cases, it is observed that the values of the elements in $J$-matrix are represented by floating-point numbers but the number of possible values ($N_v$) that each element takes is limited (limited variation).  Fig.~\ref{Fig_IFCA}\textbf{a} shows the distribution (histgram) of the values of the $J$-matrix corresponding to several representative combinatorial optimization problems, i.e., MIS, WMMBG (Weighted Maximum Matching in Bipartite Graphs) ~\cite{Tatsumura24}, and G-set~\cite{Gset}, where the elements of the $J$-matrixes take only two or three different values. The reason for the limited variation is that in the discrete optimization problems formulated as the Ising model, non-diagnal elements $J_{i\neq j}$ often represent binary (or digitalized) exclusive relationship between two decision variables ($s_i$ and $s_j$). Note that the TSP (Traveling Salesman Problem) shown in Fig.~\ref{Fig_IFCA}\textbf{a} is an exception for the limited variation, where $J_{i\neq j}$ can take continuous values corresponding to distances between two cities. See the Methods section for WMMBG, G-set, and TSP.

In this architecture, the $J_{i,j}$ is losslessly encoded using an indexed $J$-Table generated based on the distribution of $J$-matrix elements.  (Fig.~\ref{Fig_IFCA}\textbf{b}).  The indexed $J$-Table comprises the pairs of unique real values in the $J$-matrix and their corresponding integer indices $t$, and the number of the entries is equal to the number of data patterns $N_v$.  The number of bits required to represent an index value $t$ is $\lceil \log_2N_v \rceil$.  The lossless encoding is performed by replacing the original values $J_{i,j}$ of the $J$-matrix with the corresponding indexed (encorded) value $J_{i,j}^c$ in the $J$-Table.  If the original real values are represented in float32 (32-bit floating-point numbers), the size of the $J$-matrix becomes $\lceil \log_2N_v\rceil/32$. When $N_v$ is small, the data size can be compressed. For example, in the case of MIS ($N_v = 2$), the $J$ size is reduced to be $1/32$.  By table-lookup in the $J$-Table, the original values can be restored from $J_{i,j}^c$ without any loss of information.

This encoding method further provide significant benefits when combined with hardware acceleration systems (Fig.~\ref{Fig_IFCA}\textbf{c}).  A hardware acceleration system refers to a technology that accelerates specific processes (in this case, the Ising machine) using dedicated hardware (such as FPGA) instead of a general-purpose processor (CPU).  In systems where the Ising machine is implemented on a device, the Ising model is transmitted from the host, where the CPU operates, to the device prior to the execution of the Ising machine. During this process, the $J$-matrix with a large data size increases the transmission time.  Considering limited resources available for the dedicated hardware, the memory size required to store the $J$-matrix can becomes relatively large (can be a dominant limiting factor).  By encoding the $J$-matrix, both the data transmission time and the memory size can be reduced.

Furthermore, the proposed encoding method can enhance the parallelism of computation using the limited resouces.  We propose a method to efficiently implement multiply accumulate (MAC) operations with less resources using encoded states. MAC operations, which combine multiplication and addition, are frequently used in digital signal processing, machine learning, and matrix operations. The SB, adopted as the Ising machine in this study (see the Methods section for details), also execute numerous MAC operations. The MAC using the encoded states in this method can be described as follows:

\begin{equation}\label{Eq_mac}
\sum_{j=1}^N x_j J_{i,j} \Longleftrightarrow \sum_{t=1}^{N_v}\left(\text{TBL}[t] \sum_{j=1}^N x_j \chi_t(J_{i,j}^c)\right).
\end{equation}
The left-hand side represents the conventional MAC operation to calculate the many-body interactions to $i$th non-linear oscillator in SB. The right-hand side represents its transformed version, where \text{TBL}[$t$] represents the real value corresponding to index $t$ in the indexed $J$-Table, and $\chi_t(v)$ is an indicator function that returns 1 when $v=t$ and 0 otherwise.  $x_j$ is the position variable for $j$th non-linear oscillator in SB, and it is crucial that the cumulative operation of $x_j$ is factored out.  The circuit architecture (indexed $J$-specialized MAC) corresponding to the right-hand side in Eq.~(\ref{Eq_mac}) is shown in (Fig.~\ref{Fig_IFCA}\textbf{d}).  After the selective cumulative-addition of $x_j$ by an accumulator (ACC$_t$) for each index $t$, the original real value (\text{TBL}[$t$]) obtained from the indexed $J$-Table is multiplied. In other words, the selective cumulative-addition is performed while $J_{i,j}$ remains encoded.  In addition, since $x_j$ is a physical quantity with a predetermined dynamic range (in the case of bSB used here, the position variable is ($-1.0 \leq x \leq 1.0$)), a fixed-point data type (e.g., int16) can be used for ACC$_t$. Generally, fixed-point arithmetic units can be implemented with lower latency and fewer resources compared to floating-point units.  Thus, the MAC operations can be implemented with low resources when $N_v$ is sufficiently small, leading to the overall acceleration (more pallallelism) of the Ising machine.

\subsection{ML-assisted scalable and diversified architecture}\label{sec_scalable}
\begin{figure}[t]
\centering
\includegraphics[width=17.2 cm]{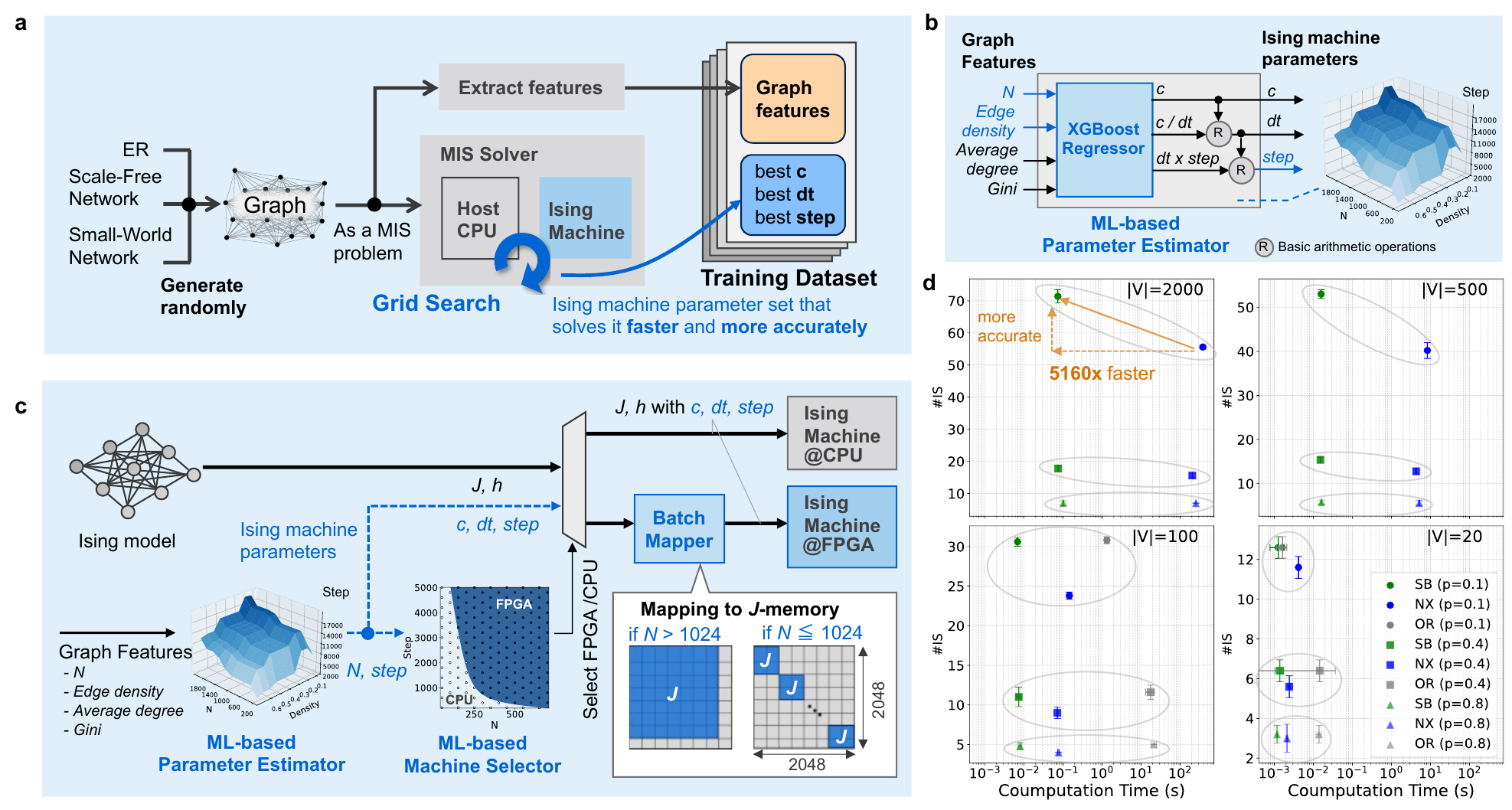}
\caption{ML-assisted scalable and diversified architecture. The MIS (maximum independent set) problem is chosen as a motif. 
(a) Preparation of the training dataset. Generate 2840 random graphs based on three different basic structures.  Using an Ising machine-based MIS solver, perform a grid search to find the Ising machine parameters that can produce the best independent set for each graph more quickly. Record the best parameters with graph features in the dataset.
(b) ML-based parameter estimator trained based on a XGBoost regression model. The 3D graph visualizes the response of $steps$ to $N$ and $density$, with the values of $average\ degree$ and $Gini$ fixed.
(c) System architecture, which comprises multiple Ising machines (FPGA/CPU-implmented), a machine selector, a parameter estimator, and a batch mapper. By using the machine selector and the parameter estimator, one of Ising machines and the control paramters for the Ising machine are selected for each graph (problem) depending on the graph features.  A batch processing mechanism to efficiently solves small problems with the fixed-size FPGA-based Ising machine is also equipped.
(d) Performance comparison of the three MIS solvers based on SB (heuristic, this work), NetworkX (heuristic), and OR-Tools (exact-solution), showing the number of independent sets
(\#IS) $\uparrow$ versus the computation time $\downarrow$ for various sizes and characteristics of graphs with edge densities $p$ (0.1, 0.4, 0.8) and node numbers $|V|$ (2000, 500, 100, 20). Each data point represents the mean value of \#IS and computation time with error bars indicating the standard deviation for five instances. The legend in the bottom-right graph applies to all four graphs.
}

\label{Fig_scalable}
\end{figure}

To handle the various sizes and characteristics of dynamically changing problems, we propose a scalable and diversified achitecutre for Ising machine-based solvers, which consists of three key methods. In this work, we choose the MIS prblem as a representative problem. 

The first method is how to construct a machine learning-based parameter estimation model.  Typically, an Ising machine has its own control parameters, which have optimal values that depend on the problems to be solved. We build and train a parameter estimation model using a large amount of training data, which enables obtaining optimal parameters with low latency at runtime. As the training dataset (Fig.~\ref{Fig_scalable}\textbf{a}), we prepare 2840 graphs with various sizes and characteristics, which are generated from the representative networks of Erd{\"o}s-R{\'e}nyi (ER)~\cite{Erdos}, Barab{\'a}si-Albert (Scale-free network, SF)~\cite{Barabashi99}, and Watts-Strogatz (Small-world network, SW)~\cite{Watts98} with randomly-selected structural parameters. The explanatory variables of the estimation model are the number of nodes, edge density, average degree, and Gini coefficient~\cite{Gini}. The objective variables of the estimation model are the three control parameters of the SB, i.e. $c$, $dt$, $step$ (see the Methods section). A grid search for those control parameter is executed with the SB-based Ising machine for each graph (each problem) to find the parameter set that solves the problem \textit{faster and more accurately}. The best parameter set found by the grid serch (the objective variables) is then recorded along with the graph features (the corresponding explanatory variables).  To evaluate the degree of \textit{faster and more accurately}, we introduce a metric called time-to-sectional-target~(TTST).  While other metrics called time-to-solution~(TTS) or time-to-target~(TTT)~\cite{King15} are often used for evaluating the performance of Ising machines~\cite{ sbm2}, these metrics require the exact solution of the problem to be known.  Since the problems (graphs) are randomly generated, the exact solutions are difficult to be obtained.  Here, the value that is 99\% of the maximum value (the maximum size of independet set) obtained within the executed grid search is defined as the sectional target~(ST). TTST is formulated as follows:

\begin{equation}\label{Eq:TTST}
\text{TTST}=
\begin{cases}
T_{\text{COM}}\frac{\log(1-0.99)}{\log(1-P_s)}, & (\text{if } P_s < 0.99),\\
T_{\text{COM}}, & (\text{if } P_s \geq 0.99),
\end{cases}
\end{equation}
where $T_{\text{COM}}$ is the computation time per shot (per execution of the Ising machine) and $P_s$ is the success probability to find solutions better
than or equal to the ST.
The parameter set that minimizes TTST is regarded as the best one by the grid search.

As the parameter estimator in this work, a regression model is constructed using XGBoost (eXtreme Gradient Boosting)~\cite{XGBoost}~(Fig.~\ref{Fig_scalable}\textbf{b}). The XGBoost is a machine learning library based on the gradient boosting algorithm and includes regularization features to prevent overfitting, which thus can realize high predictive accuracy and computational efficiency and can be applied to various tasks such as regression and classification. There are cases where different sets of explanatory variables ($c$, $dt$, $step$) yield the same TTST. This is due to the dependencies among the explanatory variables, making it impossible to learn $c$, $dt$, and $step$ independently. Therefore, we first learn $c$ independently as the reference variable, and then learn the other variables as ratios related to it ($c/dt$, $dt\times step$). Once $c$ is determined, $dt$ and $step$ are uniquely determined through simple arithmetic operations (see the Methods section for details). Overall, the parameter estimator takes as input four graph features (number of nodes ($N$), edge density, average degree, Gini coefficient) and then estimates the three Ising machine parameters ($c, dt, step$).

The second method is how to keep high computational efficiency with a fixed-size Ising machine (a hardware accelerator) when solving various sizes of problems. As shown in Fig.~\ref{Fig_scalable}\textbf{c}, the proposed architecture has a batch processing (multi-shot execution) mechanism that efficiently solves small problems with a relatively large Ising machine. The hardware Ising machine like an FPGA-based SB machine used in this work is usually designed to treat the Ising model with a predetermined maximum number of spins $N_{MAX}$ ($N_{\text{MAX}} = 2048$ in this work). The computational efficiency become the highest when solving the maximum size of Ising problem (the problem size $N$ is equal to $N_{MAX}$), while it lowers when the problem size is small ($N < N_{MAX}$) due to increased the idle time of hardware resources (e.g. the arithmetic units). In the batch processing mechanism, when the problem size $N$ is less than or equal to the half of $N_{\text{MAX}}$, the $J$ data of the problem is duplicated (i.e. the problem itself is duplicated) and assigned to the diagonal part of the $J$ memory of the Ising machine $\lfloor N_{\text{MAX}}/N \rfloor$ times with zero pudding to remaining parts (the Batch Mapper in Fig.~\ref{Fig_scalable}\textbf{c}). This enables obtaining multiple ($\lfloor N_{\text{MAX}}/N \rfloor$) different approximate solutions by one execution of the Ising machine (there is no interaction between duplicated Ising models owing to the zero pudding), effectively making the execution speed $\lfloor N_{\text{MAX}}/N \rfloor$ times faster. Generally, an Ising machine is a heuristic solver that does not guarantee the exact solution to be obtained and is capable of producing different-quality solutions by changing the initial values or other control parameters. In the case of SB, different solutions can be obtained by changing the initial values of position $x$ or momentum $y$ variables. After the obtaining different solutions, we can choose the best solution, which improves the quality of solution. 

The third method is another approch also to efficietly solve various-size problems. We prepare various-size Ising machines with different implementations (CPU-based and FPGA-based Ising machines in Fig.~\ref{Fig_scalable}\textbf{c}) and a selection mechanism based on machine learning methodology~(the Machine Selector in Fig.~\ref{Fig_scalable}\textbf{c}). Even with the batch mapping mechanism, as the problem size decreases, the execution by the FPGA-based implementaion becomes inefficient. Compared to the computation time depending on the $N$ (the problem size) and $step$~(the number of iterations in SB argorithm), the data transfer time (the time for offloading from CPU to FPGA) becomes non-negligible.  We design a software (CPU)-implemented Ising machine that operates relatively faster only for smaller problems and also prepare a \textit{machine selector} to select a faster machine depending on the situation (the features). The machine selector introduced here is a second machine learning model, distinct from the previously mentioned parameter estimator. In the machine selector, the training dataset consists of recorded system-wide latency (computation time) for each Ising machine on the host computer for multiple sets of $N$ and $step$. The goal is to perform binary classification to determine which Ising machine operates faster. We used a polynomial logistic regression model as the machine learning method. Using $N$ determined by the problem itself and $step$ determined by the parameter estimator described aboves, the model selects the fastest implementation based on these features. Thus, scalability~(agility) toward very small problems are ensured.

Following the proposed architecture, we constructed and evaluated a MIS solver in terms of computation time and solution accuracy (the size of the independent set found, \#IS) and then compared it with conventional MIS solvers (NetworkX~\cite{NetworkX}, OR-Tools~\cite{ORTOOLS}). NetworkX is an open-source Python library that includes an MIS solver based on heuristic algorithms. OR-Tools is an open-source toolkit for combinatorial optimization problems, capable of constructing an exact MIS solver by formulating it as an integer programming problem. In the evaluation, the graph sizes $|V|$ weve set to 2000, 500, 100, and 20, and the edge densities $p$ were set to 0.1, 0.4, and 0.8. For each combination of $|V|$ and $p$, five different graphs (instances with different characteristics) were randomly generated and processed by the MIS solvers.  Fig.~\ref{Fig_scalable}\textbf{d} shows the evaluation results of the proposed and conventional MIS solvers.  This computation time shows the system-wide latency including not only the execution time of Ishing machine but also all the other time components shown in Fig.~\ref{Fig_OverView}.  While the computation time of the conventional MIS solvers increases with the growth of $|V|$, the proposed MIS solver consistently achieves a computation time of approximately 100 ms regardless of $|V|$. For $|V| = 2000$~$(p = 0.1)$, the proposed MIS solver is 5160 times faster than the heuristic solver NetworkX.  OR-Tools failed to provide a solution within the timeout period of 3600 seconds for some instances with $|V| = 500$ and for all instances with $|V| = 2000$. In terms of solution accuracy, while there are cases where the proposed MIS solver slightly lags behind the exact solver OR-Tools, the proposed solver consistently outperforms NetworkX and maintains a high standard.  These results demonstrate that the proposed MIS solver has advantages in both speed and solution accuracy for problems with different sizes and characteristics.

\subsection{Demonstration}\label{sec_demo}
\begin{figure}[t]
\centering
\includegraphics[width=17.2 cm]{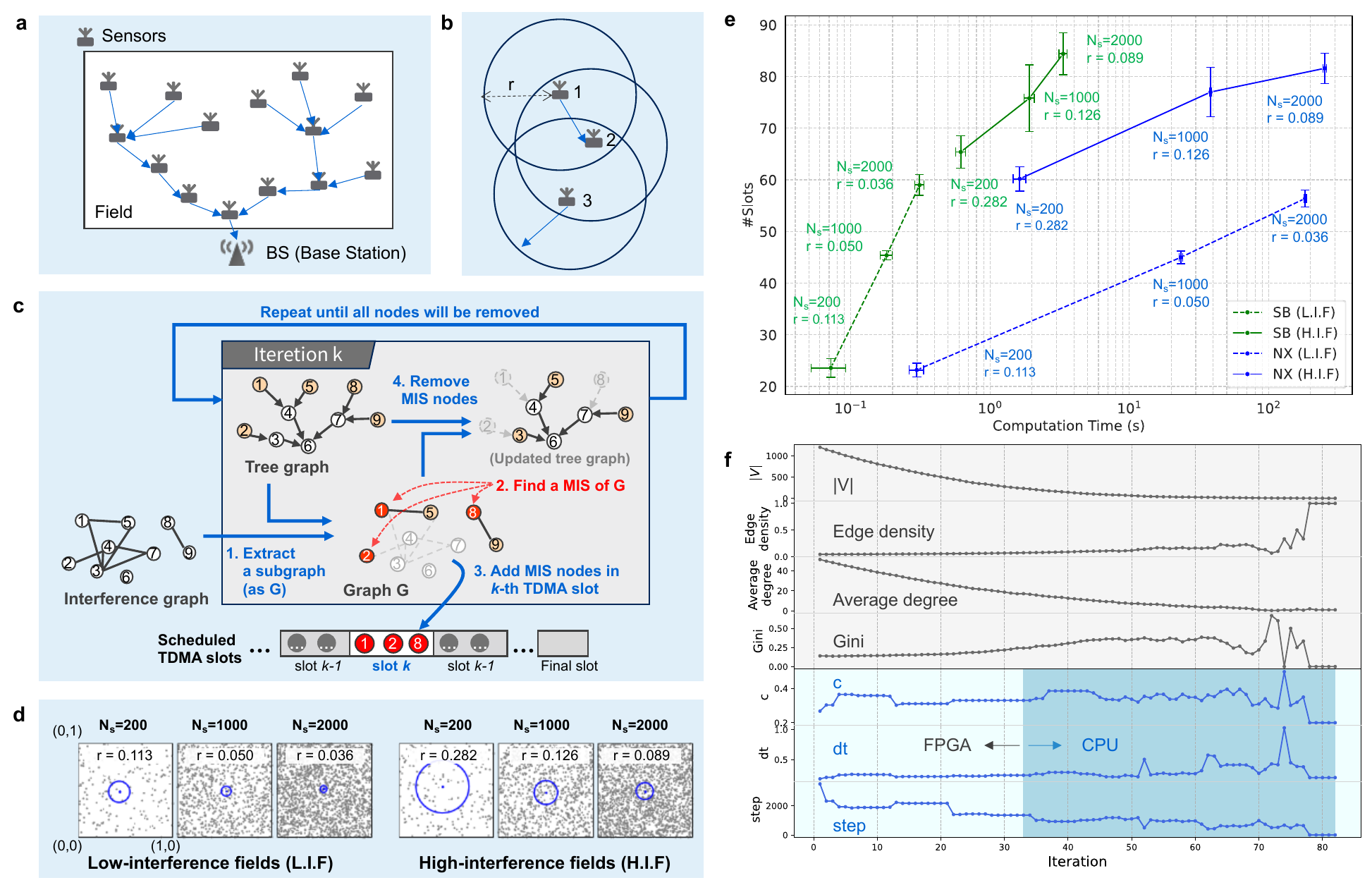}
\caption{Demonstration of TDMA scheduling in wireless multi-hop networks.
(a) An example of transmission paths with a tree topology. Each sensor sends its data to its parent node, and eventually, the data from all sensors is collected at the base station (BS).
(b) An example of interference as a constraint. Consider a situation where data is being transmitted from node 1 to node 2 and simultaneously from node 3 to another node. If node 2 in the receiving state is in the communication radius \textit{r} centerded at node 1 and node 3, the interference of the multiple signals happens.
(c) TDMA scheduling procedure. Nodes are iteratively assigned to time slots (the $k$-th TDMA slot) based on interference constraints. The tree graph shrinks as leaf nodes are removed at each iteration. The maximum independent set (MIS) of the subgraph $G$, formed by leaf nodes, is found for each time slot, and these nodes are scheduled and removed from the tree graph. This process repeats until all nodes are scheduled.
(d) Experimental setup. $N_s$ sensor nodes placed randomly in a $1.0 \times 1.0$ field, with low interference fields (L.I.F) and high interference fields (H.I.F). The communication radius $r$ varies to create different interference levels.
(e) The total number of TDMA slots (\#slots$\downarrow$) versus the computation time$\downarrow$ for simulated bifurcation (SB)-based and NetworkX (NX)-based MIS solvers when changing the number of sensor nodes ($N_s$) and the communication radius ($r$).
(f) Transition of graph features (edge density, average degree, Gini coefficient) and selected Ising machine parameters ($c$, $dt$, $step$), as well as the decision on which Ising machine to use, during iterations for $N_s = 2000$ in H.I.F.  The light-colored area indicates the FPGA-based Ising machine is selected, while the dark-colored area indicates the CPU-based Ising machine is chosen.
}
\label{Fig_demo}
\end{figure}

Combinatorial optimization problems in wireless network systems have been tackled with various classical and quantum approaches~\cite{ishizaki19, Sakaki16, Wang16, Zhou17, Charles23}. A TDMA scheduling problem to avoid interference in wireless multi-hop networks proposed by Ishizaki~\cite{ishizaki19} is chosen as a motif of dynamically changing problems to demonstrate the performance of the proposed MIS solver. We also developed an interactive application with graphical user interface for the demonstration. For a more visual understanding, see the application videos (Video1, Video2) provided as the Supplementary information.

Fig.~\ref{Fig_demo}\textbf{a} shows an example of static tree-topology network (a directed graph) which determines the transmission paths of the senser network system; the distination and source nodes (sensors) of a directed edge (an arrow) are in the relationship of parent and child nodes. A parent node has to receive data from all the child nodes in a time slot or over several time slots (receiving state) and then the node (in turn as a child node) transmits the aggregated data to its parent node in a following time slot (transmitting state). The transmission process is repeated slot-by-slot until all data is finally collected at the base station (BS). A finite range that the transmission signal reaches is defined as the communication radius $r$ (common for all the senser nodes). After the transmission graph is determined by a graph-theory method, the TDMA scheduler determines which node transmits in which time slot. Multiple nodes can transmit simultaneously unless wireless interference occurs. It is prohibited that multiple transmission signals reach a receiving node simultaneously (in a time slot) in order to avoid signal interference. For an example, in the situation shown in the Fig.~\ref{Fig_demo}\textbf{b}, the node 3 is not allowed to transmit when the node 1 is transmitting because the node 3 is in the radius $r$ centered at the node 2 that is the destination (the parent node) of the node 1. The relationships between nodes that cannot transmit simultaneously is summarized as an interference graph (Fig.~\ref{Fig_demo}\textbf{c}), where the nodes connected by an edge are not allowed to transmit simultaneously. The TDMA scheduling is the problem to determine the assignment of the nodes to time slots so as to minimize the number of slots to be needed based on the predetermined transmission graph and interference graph.

Fig.~\ref{Fig_demo}\textbf{c} illustrates the iteration procedure of the TDMA scheduling, where the $k$th iteration determines the nodes assigned to the $k$th time slot. The procedure uses two temporary graphs: a tree graph and a subgraph ($G$). The tree graph starts from the given transmission graph and gradually shrinks as a set of leaf nodes are removed at each iteration. Here, the removed leaf nodes are assigned to the time slot corresponding to the iteration. The iteration is repeated until all nodes are removed from the tree graph. At each iteration, the leaf nodes are the candidate nodes to transmit. The subgraph $G$ at each iteration comprises nodes corresponding the leaf nodes of the tree graph, where the presence or absence of edges between the nodes is determined according to the interference graph. The maximum independent set (MIS) of $G$ corresponds to the maximum number of nodes that can transmit simultaneously in the $k$th slot. We use the proposed MIS solver to find the MIS of $G$. The size and characteristics of $G$ dynamically change iteration-by-iteration as illustrated in Fig.\ref{Fig_OverView}. The figures of merit for a method for TDMA scheduling are the number of slots achieved (\#Slots$\downarrow$) and the total computation time ($\downarrow$) needed to obtain the schedule.  In this study, we adopted a simple greedy algorithm as described above, for the purpose of evaluating the scalable MIS solver. Intuitively, it is expected that higher MIS accuracy (the ability to find larger independent sets) would result in a smaller \#Slots. However, this is not always the case due to the structure of the tree network. Therefore, it is important to note that in the subsequent evaluations, \#Slots should be considered as a supplemental metric.

To compare the perfomance of the proposed MIS solver (SB) and NetworkX (NX), we systematically prepare experimental conditions (fields) as shown in Fig.~\ref{Fig_demo}\textbf{d}. The numbers of senser nodes ($N_s$) in the fields are 200, 1000, or 2000, where they are randomly placed. There are two type of fields: low interference fields (L.I.F) and high interference fields (H.I.F), where the averaged number of the neighboring nodes within the communication radius $r$ centered at a node are 8 for L.I.F and 50 for H.I.F. Note that in this work, the transmission graph is constructed based on the shortest paths from the BS to each node to create a balanced tree structure instead of the minimum spanning tree (MST) used in Ref.\cite{ishizaki19} because the MST can sometimes produce very long sequential paths.

 Fig.~\ref{Fig_demo}\textbf{e} shows the total number of slots (\#Slots$\downarrow$) versus the computation time ($\downarrow$) for the proposed MIS solver (SB) and NetworkX (NX).  In all the conditions, the proposed MIS solver demonstrates a computational speed advantage over NX, with no significant difference in \#Slots.  Fig.~\ref{Fig_demo}\textbf{f} shows, for the case of $N_s = 2000$ and H.I.F., how the graph features of the MIS problems (the subgraph $G$) change as the iteration proceeds, what Ising machine parameters  ($c, dt, step$) are selected by the parameter estimator for each iteration, and which Ising machine (FPGA-based or CPU-based) is selected by the machine selector. The graph features change in a wide dynamic range (e.g. $|V|$ decreases from 1202 to 1, while edge density increases from 0.042664 to 1.0). Responding to the variety of the problems, a wide variety of combinations of parameters and machines are appropriately selected by the ML-based technique, leading to the overall performance advantages of the proposed methodology shown in Fig.~\ref{Fig_demo}\textbf{d}.

\section{Discussion}\label{sec_discussion}
To achieve high-speed combinatorial optimization for dynamically changing problems, we propose a methodology that combines an embedded Ising machine, which is based on the quantum-inspired SB algorithm and the indexing fast-computation architecture for high-speed computation using data compression, with machine learning. We define features that characterize the problem graphs and learn the relationship between these features and the multiple interdependent Ising parameters by training on data derived from a set of randomly generated graphs. The MIS solver based on this methodology has been shown to be over 5000 times faster than conventional methods. In the practical motif of TDMA scheduling in wireless multi-hop networks, the machine learning model effectively selects parameters according to problem characteristics, demonstrating the effectiveness of the proposed methodology in terms of computational time. This research enhances the practical applicability of Ising machines and contributes to the development of an interdisciplinary field that spans quantum computing, machine learning, computer science, and discrete mathematics.

There are several directions for future work. First, further investigation is needed regarding the generalization of the design of machine learning models. Although we have demonstrated an approach using a specific algorithm to estimate the parameters of the Ising model, it has not been shown to be effective for different classes of combinatorial optimization problems. It would be necessary to verify its effectiveness in other problems and to study methods for constructing machine learning models that are common to many combinatorial optimization problems. Additionally, the application of GNNs (Graph Neural Networks) can also be considered in this context. Second, we adopted TDMA scheduling as a practical application in this study, but our methodology would be extended to various applications, including real-time applications such as finance, automotive, and robotics.
\clearpage

\section*{Methods}
\subsection*{Ising and QUBO problems}
The Ising model and QUBO~(Quadratic Unconstrained Binary Optimization) are important mathematical models for solving combinatorial optimization problems. They are closely related (one-to-one correspondence) and can be applied to many practical problems.  The Ising model is a statistical mechanics model that describes the properties of magnetic materials.  It uses spin variables $s_{i}\;(\in \{-1,1\})$ as decision variables, and the Hamiltonian of the Ising model is expressed as shown in Eq.~(\ref{IsingEnergy}).  QUBO is an optimization problem with binary variables $b_{i}\;(\in \{0,1\})$ and a quadratic cost function as follows:

\begin{equation}
H_{\mathrm{QUBO}}=\sum_{i}^{N}\sum_{j}^{N}Q_{i,j}b_{i}b_{j},
\end{equation}
where $Q$ is referred to as the QUBO matrix.  The QUBO can be transformed into the Ising model using the following conversions.

\begin{align}
s_{i}&=2b_{i}-1,
\label{eq:conv1}\\
J_{i,j}&=
\begin{cases}
-\frac{Q_{i,j}}{2} & (\text{if}\;i\neq j),\\
0 & (\text{if}\;i= j),
\end{cases}
\label{eq:conv2}\\
h_{i}&=\sum_{j}^{N}\frac{Q_{i,j}}{2}.
\label{eq:conv3}
\end{align}

The MIS problem is formulated as a QUBO in \cite{lucas14,Hidaka23} as follows,

\begin{equation}\label{Eq_mis_b}
b_i=
\begin{cases}
1, & (\text{Node } i \text{ is an element of the independent set}),\\
0, & (\text{Node } i \text{ is not an element of the independent set}).
\end{cases}
\end{equation}
\begin{equation}\label{Eq_mis}
H_{\text{MIS}}=-B\sum_{i \in V}b_i + A\sum_{(i,j)\in E}b_ib_j.
\end{equation}
The first term corresponds to an objective fucntion to be minimized (an incentive to include more nodes) and the second term corresponds to a penalty function that increases if connected (not independent) edges ($\in E$) are included in the selected subset. The coefficients $A$ and $B$ determines the relative strength of each term.  This QUBO formulation is transformed into the Ising model using Equations Eqs.~(\ref{eq:conv1})-(\ref{eq:conv3}),
resulting in Eqs.~(\ref{Eq_mis_J}) and (\ref{Eq_mis_h}).

The WMMBG (Weighted Maximum Matching in Bipartite Graphs) is also formulated as a QUBO~\cite{Tatsumura24} as follows:

\begin{equation}\label{Eq_wmmbg_b}
b_{l,r}=
\begin{cases}
1, & (\text{if node $l$ and node $r$ are matched}),\\
0, & (\text{if node $l$ and node $r$ are unmatched}).
\end{cases}
\end{equation}
\begin{equation} \label{Eq_wmmbg}
		H_{\text{WMMBG}} = - \sum_{l=1}^n\sum_{r=1}^n W_{l,r}b_{l,r} +
		C\left( \sum\limits_{l=1}^n \Bigl( \sum\limits_{r=1}^n b_{l,r}-1 \Bigr) ^2 +
		\sum\limits_{r=1}^n \Bigl( \sum\limits_{l=1}^n b_{l,r}-1 \Bigr) ^2 \right),
\end{equation}
where $n$ is the number of nodes in an independent set (assuming for simplicity that the number of nodes in the two independent sets is the same), $C$ is a positive constant representing the constraint weight and $W_{l,r}$ represents the weight between node $l$ and node $r$.  The distribution of the $J$ when converted to the Ising model is shown in Fig.~\ref{Fig_IFCA}\textbf{a}.

G1 and G6 in Fig.~\ref{Fig_IFCA}\textbf{a} are the instances of the well-known G-set~\cite{Gset}, which serves as a benchmark for the MAX-CUT problem. The MAX-CUT involves partitioning the set of vertices of a graph into two subsets such that the sum of the weights of the edges cut by this partition is maximized. In the MAX-CUT, each vertex $i$ of the graph is assigned an Ising spin ($s_i=\pm 1$) whose sign represents a subset. The Hamiltonian of MAX-CUT is  defined as follows,

\begin{equation}\label{Eq_maxcut}
H_{\text{MAX-CUT}}=-\frac{1}{2}\sum_{(i,j) \in E} w_{i,j}(1-s_i s_j),
\end{equation}
where $E$ is the set of edges, and $w_{i,j}$ is the weight of the edge. In G1, the edge weights are binary, taking values of either -1 or 0, while in G6, the edge weights are ternary, taking values of -1, 0, or 1. Therefore, the possible values of the $J$-matrix are similarly limited.

The instance gr24 in Fig.~\ref{Fig_IFCA}\textbf{a} is from the TSPLIB~\cite{TSPLIB}, which serves as a benchmark for the TSP (Traveling Salesman Problem). TSP is a well-known example of a combinatorial optimization problem. Given the costs between all pairs of cities, it involves finding the route with the minimum total cost that visits each of the $n$ cities exactly once. The TSP can be formulated as a QUBO problem~\cite{lucas14}. Let $b_{i,j}$ be a binary variable that equals to 1 if city $i$ is visited at the $t$-th position, and 0 otherwise, the Hamiltonian of TSP is defined as follows,

\begin{equation}\label{Eq_tsp}
H_{\text{TSP}}=\sum_{t=1}^n\sum_{i=1}^n\sum_{j=1}^nd_{i,j}b_{t,i}b_{t+1,j}+C\left(\sum_{t=1}^n\left(\sum_{i=1}^nb_{t,i}-1\right)^2+\sum_{i=1}^n\left(\sum_{t=1}^nb_{t,i}-1\right)^2\right),
\end{equation}
where $d_{i,j}$ represents the distance between cities $i$ and $j$, $b_{n+1,i}=b_{1,i}$, and $C$ is a positive constant representing the constraint weight.

\subsection*{Simulated bifurcation}

Simulated bifurcation (SB)~\cite{sbm1,sbm2} is a quantum-inspired~\cite{sbm1,Qinspired24}, highly-parallelizable~\cite{FPL19,NatEle,Kashimata24}, metaheuristic algorithm for computationally-hard combinatorial (or discrete) optimization.  SB-based Ising machine belongs to a group of oscillator-based Ising machines~\cite{honjo21,kalinin20,PoorCIM19,moy22,albertsson21,wang21, SimCIM21,Graber24}.  The SB finds the optimal (exact) or near-optimal solution of the Ising problem by simulating the time-evolution process of coupled nonlinear oscillators according to the Hamilton's equations of motion (without energy-dissipative or noise-based mechanisms).  The SB has several variants including adiabatic SB, ballistic SB, and discrete SB, which differ in terms of nonlinearity~\cite{Nonlinearity21} anddiscreteness~\cite{sbm2}. 

In the SB, the $i$th nonlinear oscillator corresponds to the $i$th Ising spin and its state is described by the position and momentum ($x_i$, $y_i$). The update procedure of $x_i$ and $y_i$ for the ballistic SB, used in this work, is as follows~\cite{sbm2}.

\begin{align}
y_i^{t_{k\!+\!1}} &\gets y_i^{t_k} + \left[-(a_0-a^{t_k})x_i^{t_k} -\eta h_i + c_0\sum_{j}^{N}J_{i,j}x_j^{t_k}\right]\Delta_t,
\label{eq:y.bSB}\\
x_i^{t_{k\!+\!1}} &\gets x_i^{t_k} + a_0 y_i^{t_{k\!+\!1}}\Delta_t,
\label{eq:x.bSB}\\
(x_{i}^{t_{k\!+\!1}}, y_{i}^{t_{k\!+\!1}})&\gets
\left\{ 
\begin{alignedat}{2} 
(\mathrm{sgn}(x_{i}^{t_{k\!+\!1}}), 0) & \;\;(\text{if}\;|x_{i}^{t_{k\!+\!1}}|>1), \\
(x_{i}^{t_{k\!+\!1}}, y_{i}^{t_{k\!+\!1}}) & \;\;(\text{if}\;|x_{i}^{t_{k\!+\!1}}|\le1),
\end{alignedat} 
\right.
\label{eq:wall}
\end{align}
where $a_0$, $c_0$ and $\eta$ are positive constants, $a^{t_k}$ is a control parameter increasing from zero to $a_0$, and $\mathrm{sgn}(x) (=\pm 1)$ is the sign function.  Eq.~(\ref{eq:wall}) is a nonlinear transfer function~\cite{Nonlinearity21}, physically corresponding to a perfectly inelastic wallexisting at $x=\pm 1$.  The time increment is denoted as $\Delta_t$, and thus, $t_{k+1}=t_k+\Delta_t$.  After iterating the update procedure for the predetermined time steps ($N_{\rm step}$), the $i$th position $x_{i}$ is digitized to be the $i$th spin ($\pm 1$) by taking the sign of $x_{i}$.  In this work, $c_0 = \eta$ is denoted as $c$, $\Delta_t$ as $dt$, $N_{\text{step}}$ as $step$, and $a_0$ is set to 1. The values of $c, dt, step$ were estimated for each problem using the proposed parameter estimator with the assistance of machine learning.

The ballistic SB allows us to obtain better-quality solutions faster than the simulated annealing (SA) algorithm for academic benchmark problems~\cite{sbm2} and practical problems~\cite{Matsumoto22,ISCAS20,ACCESS23a,ACCESS23b,Hidaka23}.

\subsection*{Indexing fast-computing architecture}
This section describes the steps of a generalized indexing fast-computing architecture that is not limited to the MIS problem.  Algorithm~\ref{algo:gen_idx_table} shows the procedure for generating an indexed $J$-Table for any $J$-matrix.  Algorithm~\ref{algo:idx_encoding} describes the procedure for encoding a $J$-matrix with elements such as floating-point numbers using the indexed $J$-Table.  Algorithm~\ref{algo:idx_mac} presents a special MAC operation procedure for a variable $x$ with a known dynamic range (representable by fixed-point numbers) and the encoded $J^c$.  By using fixed-point numbers with a small bit width for the register and the adder of the ACC (accumulator), the resource amount for hardware implementation can be reduced.

\scriptsize
\begin{algorithm}[h]
\caption{Indexed $J$-Table generation}\label{algo:gen_idx_table}
\KwIn{The $J$-matrix of the Ising model, The number of Ising spins $N$.}
\KwOut{The indexed $J$-Table (TBL).}
$S \leftarrow \emptyset$\\
\For{$i \leftarrow 1$ \textbf{to} $N$} {
	\For{$j \leftarrow 1$ \textbf{to} $N$} {
		$S \leftarrow S \cup \{J_{i,j}\}$
	}
}
TBL $\leftarrow$ new array$[1..|S|]$\\
\For{$i \leftarrow 1$ \textbf{to} $|S|$} {
	TBL[$i$] $\leftarrow S_i$
}
\Return{$\mathrm{TBL}$}
\end{algorithm}
\normalsize

\scriptsize
\begin{algorithm}[h]
\caption{Index encoding}\label{algo:idx_encoding}
\KwIn{The $J$-matrix of the Ising model, The number of Ising spins $N$, and the indexed $J$-Table (TBL).}
\KwOut{The $J$-matrix encoded as $J^c$.}
$N_v \leftarrow \text{length}$(TBL)\\
$B$ $\leftarrow \lceil \log_2{N_v}\rceil$\\
Let $J^c$ be an array [1..$N$, 1..$N$] of $B$-bit integer\\

\tcp{\small Create a reverse dictionaly ReverseTBL for TBL.}
ReverseTBL $\leftarrow \{\}$\\
\For{$t \leftarrow 1$ \textbf{to} $N_v$} {
  ReverseTBL[TBL[$t$]] $\leftarrow t$
}
\tcp{\small Replace values with indexes.}
\For{$i \leftarrow 1$ \textbf{to} $N$} {
	\For{$j \leftarrow 1$ \textbf{to} $N$} {
    $J^c[i,j] \leftarrow$ ReverseTBL[$J[i,j]$]
	}
}
\Return{$J^c$}
\end{algorithm}
\normalsize

\scriptsize
\begin{algorithm}[h]
\caption{Indexed $J$-specialized MAC}\label{algo:idx_mac}
\KwIn{The variables $x$ defined in fixed-point notation, The $i$-th row of the matrix $J^c$ as $J^c_i$,
The number of Ising spins $N$, and the indexed $J$-Table (TBL).}
\KwOut{The result of the MAC operation as $\Delta y_i$}
$N_v \leftarrow \text{length}$(TBL)\\
$\Delta y_i \leftarrow 0$\\

\For{$t \leftarrow 1$ \textbf{to} $N_v$} {
  ACC $\leftarrow$ 0  \tcp{\small ACC is a fixed-point number.}
  \For{$j \leftarrow 1$ \textbf{to} $N$} {
    \If{$t = J^c_i[j]$}{
      $\text{ACC} \leftarrow \text{ACC} + x[j]$
    }
  }
  $\Delta y_i \leftarrow \Delta y_i + \text{convert\_to\_float(ACC) } \times$ TBL[$t$] \\
}

\Return{$\Delta y_i$}
\end{algorithm}
\normalsize

\subsection*{FPGA implementation of the Ising machine}
\begin{figure}[t]
\centering
\includegraphics[width=17.2 cm]{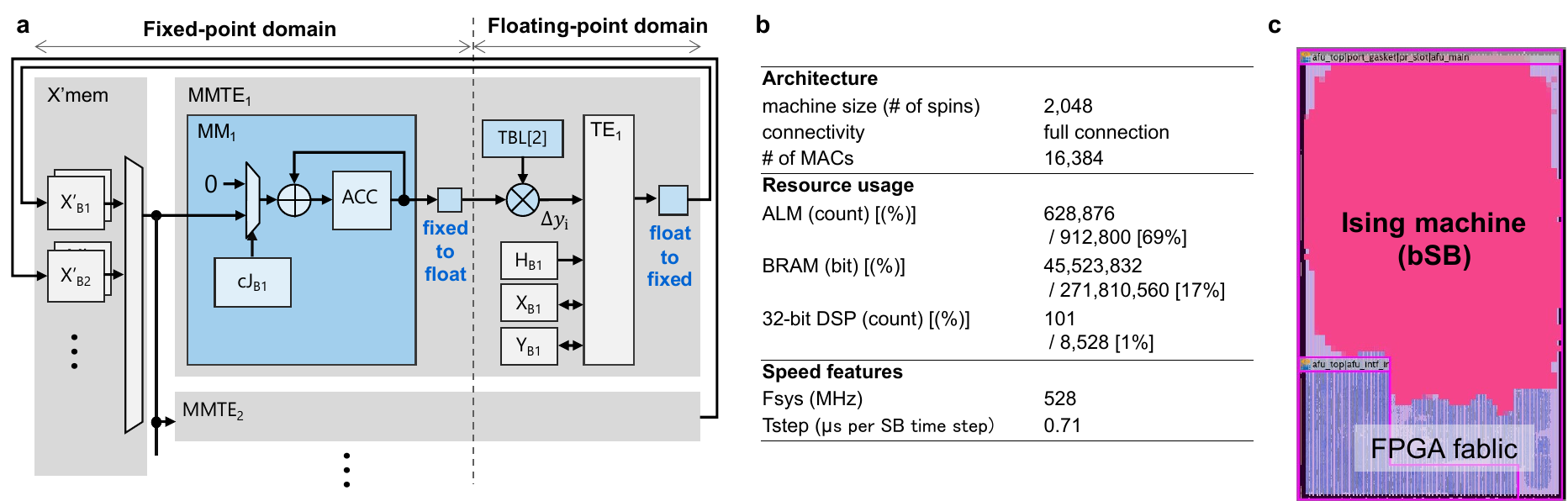}
\caption{FPGA implementation of the Ising machine.
(a) Circuit architecture of the bSB using indexing fast-computing ($N_v=2$, with TBL[1] reserved as 0).
(b) Implementation details of the SB-based Ising machine.
(c) Placement of the Ising machine in the FPGA.
}
\label{Fig_FPGA}
\end{figure}

Fig.~\ref{Fig_FPGA} shows the circuit architecture of the bSB using the indexing fast-computing architecture and the implementation results on the FPGA.  Fig.~\ref{Fig_FPGA}\textbf{a} is the circuit architecture of the bSB based on \cite{Hidaka23}.  In this study, we use the indexing fast-computing architecture with $N_v = 2$, fixing TBL[1] to 0 and allowing TBL[2] to specify any floating-point number.  The difference from \cite{Hidaka23} (shown in blue) is that the MAC operation in the MM (matrix-vector multiplication) module is replaced with the indexed-J specialized MAC in Fig.~\ref{Fig_FPGA}\textbf{d} The TE module, which performs the time evolution operation, remains unchanged from \cite{Hidaka23} and operates with floating-point numbers.  Therefore, type converter modules (fixed to float, float to fixed) are inserted before and after the TE.

Fig.~\ref{Fig_FPGA}\textbf{b} and ~\ref{Fig_FPGA}\textbf{c} show the specification details and the implementation results on the target FPGA.  We targeted the Agilex7F AGF027 on the PCIe card-type FPGA accelerator IA-840f and implemented a fully connected Ising machine with 2,048 spins.  Notably, by reducing the resources per MAC, we were able to relatively increase the number of parallel MACs to 16,384.  Another key point is that the indexing fast-computing architecture results in low memory (BRAM) usage, which means there is room to further increase the number of spins. Similarly, it is evident that the usege of DSP is also low.

\subsection*{Dependencies among the SB parameters}
\begin{figure}[t]
\centering
\includegraphics[width=17.2 cm]{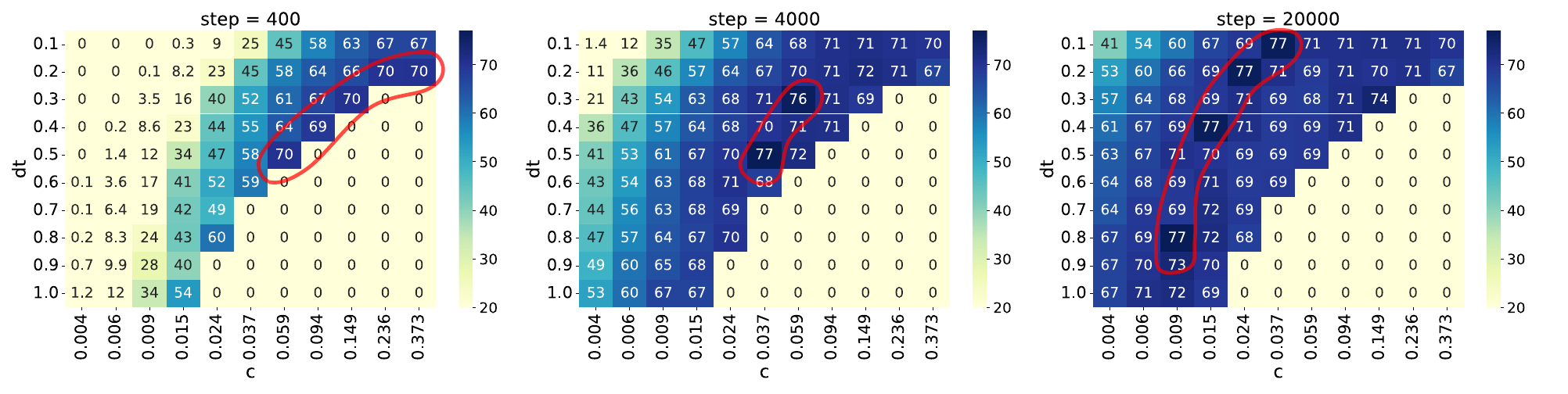}
\caption{Averaged numbers of independent sets over 100 times~(Ave.\#IS) obtained from grid search for a MIS problem (ER, $N$=2000, $p$=0.1). Heatmaps are shown for each $step$ value (400, 4000, 20000). Each heatmap represents the Ave.\#IS for the combinations of $c$ and $dt$. The red-bounded bands indicate the combinations of $c$ and $dt$ that are expected to yield the best results on the heatmap.
}
\label{Fig_params_heatmap}
\end{figure}
The parameter estimator estimates the SB parameters $c$, $dt$, and $step$ to realize minimum TTST. When training the estimator, we must pay attention to that there are dependencies among these parameters. The heatmaps in Fig.\ref{Fig_params_heatmap} show the solution quality when changing combinations of $c$, $dt$, and $step$ for a MIS problem (ER, $N$=2000, $p$=0.1), where the solution quality is the averaged number of independent sets over 100 times~(Ave.\#IS), and a higher number means a better solutions. While TTST is the evaluation metric for selecting $c$, $dt$, and $step$ as shown in Results section, we present Ave.\#IS here to clearly illustrate how the performance of the MIS solver varies with different parameters on a landscape.

It is evident from Fig.\ref{Fig_params_heatmap} that there is not just one combination of $c$ and $dt$ that yields good results (high Ave.\#IS), but rather good combinations are distributed in bands (red-bounded bands). This means that the same-quality Ave.\#IS can be obtained from different sets of $c$ and $dt$. Hence, it is not appropriate to independently learn $c$ and $dt$ to achieve good results. Here, we denote the Ave.\#IS obtained by the values of $c$ and $dt$ as $f(c, dt)$. For an example when $step=400$, two  sets of $c$ and $dt$, (0.373, 0.2) and  (0.059, 0.5), result in the same Ave.\#IS of 70, i.e. $f(0.373, 0.2)=f(0.059, 0.5)=70$, but when exchanging only $dt$ between these two sets, the resultant Ave.\#IS are totally different (significantly degraded), i.e. $f(0.373, 0.5)=0$ and $f(0.059, 0.2)=58$. Such a dependency stands also for the relationship between $dt$ (or $c$) and $step$ as seen as the shift in the position of the red-bounded bands when increasing $step$ in the three graphs of Fig.\ref{Fig_params_heatmap}.

The training data for the parameter estimator is the records of combinations of (N, edge density, average degree, Gini coefficient, $c$, $dt$, $step$) that minimize TTST for individual MIS problems. If there are multiple parameter sets that minimize TTST for a single problem, all such sets are included in the dataset. Considering the inter-parameter dependencies mentioned above, only $c$ is learned independently, while $dt$ and $step$ are not learned independently. Instead, the ratio of $c/dt$ and $dt\times step$ are learned as the target variables. Overall the explanatory variables of the XGBoost regressor model that constitutes the parameter estimator are (N, edge density, average degree, Gini coefficient), and the target variables are ($c$, $c/dt$, $dt \times step$). The parameter set of $dt$ and $step$ can be uniquely determined by $c$, $c/dt$ and $dt\times step$ estimated by the parameter estimator.

\subsection*{Experimental setup}
All experiments in this study were conducted on a single host computer (CPU: AMD Ryzen 5 5600X 6-Core Processor [3.7GHz], DDR-DRAM: 32GiB, OS: CentOS Stream 8). For the comparison with the MIS solver, NetworkX version 2.5 and OR-Tools version 9.9.3963 were used as libraries on python 3.10.10.

In Fig.~\ref{Fig_scalable}\textbf{d}, for each input setting of the MIS problem, five instances (graphs) based on different random numbers were generated, and the time taken to find the solution and the average and standard deviation of the solutions (independent set sizes) were plotted.  Similarly, in Fig.~\ref{Fig_demo}\textbf{e}, for each condition, five instances (sensor node distribution patterns) based on different random numbers were generated, and the time taken to find the solution and the average and standard deviation of the solutions (total number of slots) were plotted.

For the MIS solvers of the proposed method, NetworkX, and OR-Tools, the system-wide latency (from graph input to solution output) was used as the computation time, as shown in Fig.~\ref{Fig_OverView}. Therefore, in the case of the proposed method, the time taken for Ising model generation and parameter estimation is also included. Additionally, the proposed method performs multi-shot execution, with the number of shots set to 4.

For the MIS problem, if the input graph is fully connected, there is no solution, and if there are no edges, all nodes are independent sets.  Therefore, when applying it to TDMA scheduling in wireless sensor networks, such cases are treated specially without passing through the MIS solver.  In other words, in the former case, one node is arbitrarily selected, and in the latter case, all nodes are output as the solution.

\section*{Data availability}
The authors declare that all relevant data are included in the manuscript.  Additional data are available from the corresponding author upon reasonable request.


\section*{Acknowledgments}
The authors would like to thank Ryo Hidaka and Hayato Goto for the fruitful discussion and their support.

\section*{Competing interests}
Y.H., and K.T. are included in inventors on three U.S. patent applications related to this work filed by the Toshiba Corporation (no. 17/249353, filed 20 February 2020; no. 18/760332, filed 1 July 2024; no. 18/763618, filed 3 July 2024). The authors declare that they have no other competing interests.

\section*{Author contributions}
All the authors contributed to the whole aspects of this work, with each making the following major contribution. Y.H. developed the workflow, conceived the indexing fast-computation architecture and the ML-assisted scalable and diversified architecture, and carried out the FPGA implementation, system implementation, and evaluation of the system, as well as the implementation and evaluation of the TDMA application. T.K. and M.Y. set up the FPGA development and experimental environments. K.T. managed the project. Y.H. and K.T. wrote the manuscript.

\section*{Additional information}
\subsection*{Supplementary information}
The online version contains supplementary materials.
\begin{itemize}
\item Supplementary information 1: A document that explains the Supplementary information 1 to 3. 
\item Supplementary information 2: Video1 : Demonstration of the TDMA scheduling in wireless mutlti-hop networks (navigation mode).
\item Supplementary information 3: Video2 : Demonstration of the TDMA scheduling in wireless mutlti-hop networks (benthmark mode).
\end{itemize}

\bibliography{TDMA}
\bibliographystyle{IEEEtran}

\clearpage
\section*{Supplementary Information 1}
\subsection*{Supplementary information 1}
This document, which explains the Supplementary information 1 to 3.

\subsection*{Supplementary information 2}
\begin{figure}[h]
\centering
\includegraphics[width=14cm]{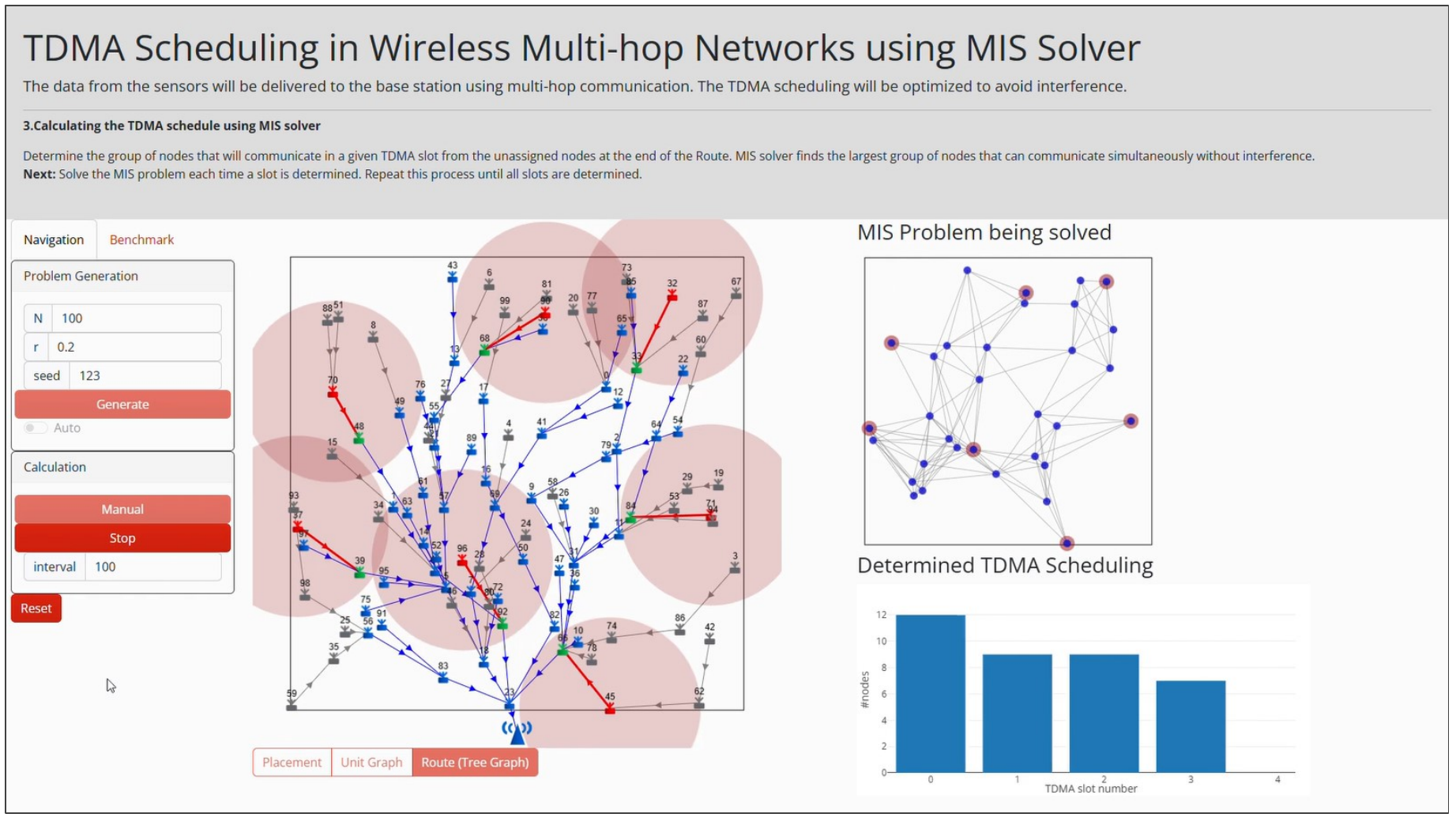}
\caption{A snapshot from the Supplementary information 2}
\label{Fig_Supp2}
\end{figure}

A video showing the operation of the TDMA scheduling demo application for wireless multi-hop networks,
navigating through the process of determining the scheduling step-by-step. A snapshot from the video is shown in Fig.\ref{Fig_Supp2}.
\begin{description}
  \item[From 0 to 15 seconds:] Starting with the number of wireless nodes ($N = 100$) and communication radius ($r = 0.2$), the placement of each node is randomly determined by triggering the ‘Generate’ button. Subsequently, the ‘Placement’ (distribution of nodes), ‘Unit Graph’ (graph connecting nodes that can communicate with each other via edges), and ‘Route (Tree Graph)’ (tree graph determined by a certain graph algorithm) are displayed sequentially. Simultaneously with the display of the tree graph, an interaction graph representing the relationships between nodes that cannot communicate simultaneously in the same time slot is also shown.
  \item[From 15 to 25 seconds:] By triggering the ‘Manual’ button, the nodes that can communicate simultaneously in the first iteration (time slot) are calculated. The calculation results are immediately displayed graphically. The solved MIS problem and the obtained solution (displayed in red) are shown in the upper right ‘MIS Problem being solved’, and the number of nodes communicating in this time slot is displayed in the lower right ‘Determined TDMA Scheduling’. On the main panel in the center, the nodes that are communicating are shown in red, and their communication range is shown in light red. Nodes in a receiving state are shown in green. It can be confirmed that the circles indicating the communication range do not overlap for nodes in the receiving state. By pressing the ‘Manual’ button several times, the calculation results for the subsequent iterations are displayed.
  \item[From 25 to 38 seconds:] For the remaining iterations, pressing the ‘Auto’ button allows the calculations and display of results to be performed automatically. After the TDMA scheduling calculations are completed, it can be seen from the upper navigation area that the calculation time was 0.037 seconds and the total number of slots was 24.
\end{description}

\subsection*{Supplementary information 3}
\begin{figure}[h]
\centering
\includegraphics[width=14cm]{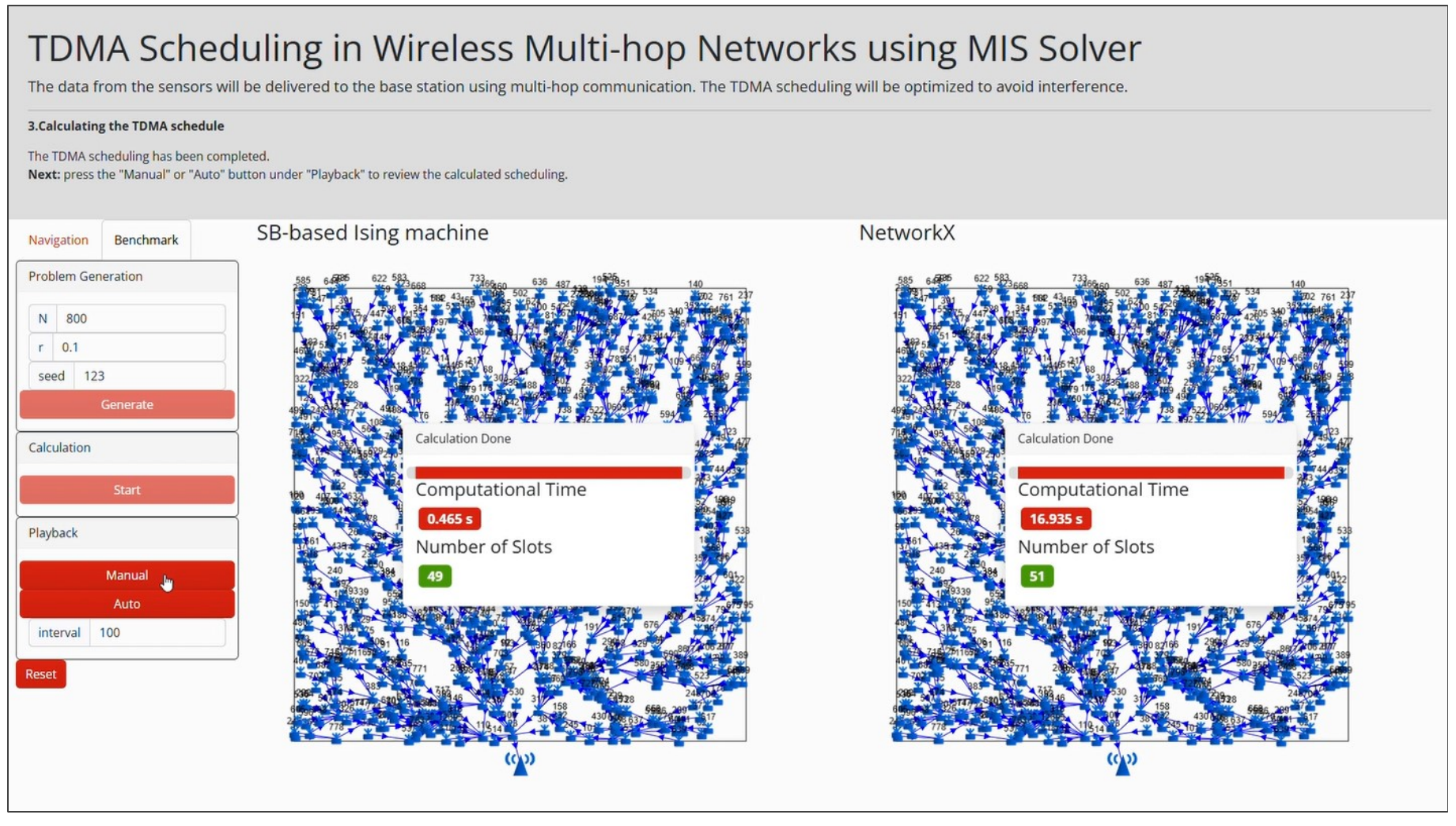}
\caption{A snapshot from the Supplementary information 3}
\label{Fig_Supp3}
\end{figure}

A video showing the operation of the TDMA scheduling demo application for wireless multi-hop networks,
benchmarking the performance differences using the MIS solver. A snapshot from the video is shown in Fig.\ref{Fig_Supp3}.
\begin{description}
  \item[From 0 to 14 seconds:] Set the application to benchmark mode and generate the problem by triggering the ‘Generate’ button with the number of wireless nodes ($N = 800$) and communication radius ($r = 0.1$). In benchmark mode, the generation up to the tree graph is performed automatically, displaying the same generated tree graph in the left ‘SB-based Ising machine’ panel and the right ‘NetworkX’ panel.
  \item[From 14 to 38 seconds:] Pressing the ‘Start’ button initiates the TDMA scheduling calculations. These calculations are performed simultaneously in separate processes for both SB-based Ising machine and NetworkX. The progress of the calculations is shown with a progress bar, and upon completion, the computational time and number of slots are overlaid on their respective panels. Here, it can be seen that SB-based Ising machine has a computational time of 0.465 seconds (with 49 slots), while NetworkX has a computational time of 16.935 seconds (with 51 slots).
  \item[From 38 to 45 seconds:] Pressing the ‘Manual’ button starts the playback of the calculation results. Similar to the navigation mode, the transmitting nodes are shown in red with their transmission range in light red, and the receiving nodes are shown in green. It can be observed that in the large network, many transmitting nodes are selected without any overlap of transmission ranges on the receiving nodes.
  \item[From 45 seconds to 1 minute and 23 seconds:] Pressing the ‘Auto’ button automatically plays back the calculation results for the remaining iterations.
\end{description}

\end{document}